\documentclass[paper]{JHEP3}
\usepackage[dvips]{graphicx}
\usepackage{epsfig,multicol,cite}

\newcommand\fverb{\setbox\pippobox=\hbox\bgroup\verb}
\newcommand\fverbdo{\egroup\medskip\noindent%
                        \fbox{\unhbox\pippobox}\ }
\newcommand\fverbit{\egroup\item[\fbox{\unhbox\pippobox}]}
\newbox\pippobox

\title{Extended $QCD_2$ from dimensional projection of $QCD_4$}

\author{Jorge Alfaro $^1$, Alexander A. Andrianov $^2$, Pedro Labra\~{n}a $^1$\\
$^1$ Facultad de F\'{\i}sica, Pontificia Universidad Cat\'olica de
Chile, Vicu\~{n}a Mackena 4860 Macul, Santiago de Chile,
Casilla 306, Chile\\
$^2$ V. A. Fock Department of Theoretical Physics, St. Petersburg
State University, 1, ul. Ulianovskaya, St. Petersburg 198504,
Russia and Istituto Nazionale di Fisica Nucleare, Sezione di
Bologna via Irnerio 46, Bologna 40126, Italy \\ \\E-Mail:
\email{jalfaro@puc.cl}, \email{andrianov@bo.infn.it} and
\email{plabrana@puc.cl}}

\abstract{
We study an extended QCD model in $(1+1)$ dimensions obtained from
QCD in $4D$ by compactifying two spatial dimensions and projecting onto the
zero-mode subspace. We work out
this model in the large $N_c$ limit and using light cone gauge but keeping
the equal-time quantization.
This system is found to induce a dynamical mass for transverse
gluons -- adjoint scalars in $QCD_2$, and to undergo a chiral
symmetry breaking with the full quark propagators yielding
non-tachyonic, dynamical quark masses, even in the chiral limit.
We study quark-antiquark bound states  which can be classified in
this model by their properties under Lorentz transformations inherited from
$4D$. The scalar and pseudoscalar sectors of the theory are
examined and in the  chiral limit
a massless ground state for pseudoscalars is revealed with a
wave function
generalizing the so called \mbox{'t Hooft} pion solution.}


\preprint{HEP-TH/0309110}
\begin{document}

\section{Introduction}

It is fairly well accepted that Quantum Chromodynamics (QCD) is a
correct theory of strong interactions giving a good description of
the high energy scattering of hadrons. However our ability to
develop its predictions at low energies, the non-perturbative
regime of QCD, is not as good. The latter problem
can be handled more easily if we consider a lower dimensional
space. In this way, QCD in two dimensions, $QCD_2$, first
studied by \mbox{'t Hooft} in the large $N_c$ limit \cite{thooft0,thooft1},
has proved to be a very interesting laboratory for
checking non-perturbative aspects of QCD.

This model has been a subject of extensive investigation,
covering the study of quark confinement,
hadron form
factors, deep-inelastic  \cite{gross1,einhorn} and hadron-hadron
scattering \cite{ellis1}, the role of instantons and solitons
\cite{grossellis, bassetto}, multi-flavor states \cite{frishman1},
quark condensation \cite{zhitnitsky} and heavy quark
physics \cite{ural}. The bosonization of $QCD_2$ was used as a tool for
the description of meson
and baryon spectra \cite{steinhardt,affleck,vento1} and of chiral
symmetry breaking \cite{affleck,vento2}.

The remarkable feature of this model is that it exhibits
quark confinement with an approximately linear Regge trajectory
for the masses of $q{\bar q}$ bound states \cite{thooft1}. As well
$QCD_2$ in the large $N_c$ limit has an interpretation in terms of
string variables \cite{gross} sharing thereby important
properties that we expect to find for real QCD in four
dimensions.

However $QCD_2$ is too simple to describe certain realistic aspect
of the higher dimensional theory. For instance, there are no
physical gluons in two dimensions and both angular momentum and
spin are absent in a $2D$ theory. Moreover in $QCD_2$ there is a
problem of  tachyonic properties of \mbox{'t Hooft} quarks
\cite{thooft1} and a realistic chiral symmetry breaking is absent
which manifests itself in the decoupling of massless pions \cite{krauth}.
Thus one needs more information from extra dimensions in order to
have a more realistic picture of QCD by using a $2D$ model. For
instance, the inclusion of boson matter in the adjoint gauge group
representation \cite{adjoint,light} may provide an information of
transverse degrees of freedom, characteristic of a gauge theory in
higher dimensions, and give a more adequate picture of strong
interaction.

In this paper we study a QCD reduced model in $(1+1)$ dimensions
which can be formally obtained from QCD in $4D$ by means of a classical
dimensional reduction from $4D$ to $2D$ and neglecting heavy K-K
(Kaluza-Klein) states. Thus only zero-modes in the harmonic expansion in
compactified dimensions
are retained. As a consequence, we obtain a two
dimensional model with some resemblances of the real theory in higher
dimension, that is, in a natural way adding boson matter in the adjoint
representation to $QCD_2$. The latter fields being scalars in $2D$
reproduce transverse gluon effects. Furthermore this model has a
richer spinor structure than just $QCD_2$ giving a better
resolution of scalar and vector states which can be classified by
their properties inherited from  $4D$ Lorentz transformations. The model is
analyzed in the light cone gauge and using large $N_c$ limit. The
contributions of the extra dimensions are controlled by the
radiatively induced masses of the scalar gluons as they carry a
piece of information of transverse degrees of freedom. We consider
their masses as large parameters in our approximations yet being much less
than the first massive K-K excitation. However this model is treated on its
own ground without any further comparison with possible contributions
of heavier K-K boson states. As well we do not
pay any attention to the mass generation via Hosotani mechanism \cite{hoso}
as it has no significance
in the planar limit of two-dimensional QCD models under
consideration, just being saturated by fermion loops sub-leading in the
$1/N_c$ expansion.

This model might give more insights into the chiral
symmetry breaking regime of $QCD_4$. Namely, we are going to show
that the inclusion of  solely lightest K-K boson modes catalyze the
generation of quark dynamical mass and allows us  to overcome the
problem of tachyonic quarks present in $QCD_2$.

The paper is written as follows, in Sect. \ref{C1} we make a formal
design of the $2D$ reduced model from QCD in $4D$ by using
Kaluza-Klein compactification and further projection onto zero-mode
 subspace. First we work with the gluon part
of $QCD_4$ which is mapped into a $2D$ $SU\!(N)$ gauge field and
two scalar fields in the adjoint representation of the reduced
gauge model. In principle, these scalars are massless but we
recover an infrared generation of masses for these scalar gluons.
Next we examine
the fermion part of $QCD_4$ and after compactifying and projecting
we obtain two
spinors minimally coupled to the $2D$ gauge field and interacting
with the scalar gluons. The full propagator for these fermions is
calculated in the large $N_c$ limit by using the Schwinger-Dyson
equation and a non-perturbative solution for the self-energy is
found  that gives a real net quark mass, even in the chiral limit
\cite{footnote1}. In Sect. \ref{C2mesons} we study bound states of
quark and antiquark by using the Bethe-Salpeter equation in the
large $N_c$ and the ladder approximations. $q{\bar q}$ states of
the model are  identified in terms of their custodial symmetries from
$4D$ Lorentz group. We focus our attention on the scalar and
pseudoscalar sectors of the theory and in Sect. \ref{C3integral}
the low mass spectrum of the scalar and pseudoscalar sector is
examined. In the chiral limit, a massless ground state is revealed
which is interpreted as the pion of the model, a projection of the
real Goldstone boson in $4D$ to  $2D$ treated in the
large-$N_c$ limit. We summarize our points and
give conclusions in Sect. \ref{C4conclusion}.

\section{Compactification of $QCD_4$ to the Reduced Model}
\label{C1}

\subsection{The Gluon Part}

We start with the $QCD_{4}$ action for gluons in $(3+1)$
dimensions:

\begin{equation}
S_{QCD}=-\int d^4x \frac{1}{2{\tilde g}^2}tr(G_{\mu\nu}^2)\, ,
\end{equation}
where:

\begin{equation}
G_{\mu \, \nu}= \partial_\mu G_\nu-\partial_\nu G_\mu-i
\left[G_\mu , G_\nu \right],
\end{equation}
with $G_\mu = G^{a}_{\mu} \,\lambda_a$. Generators $\lambda_{a}$
and structure constants $f_{abc}$ satisfy:
\begin{eqnarray}
tr(\lambda_a\lambda_b)=\frac{1}{2}\delta_{ab}\,, &
\left[\lambda_a, \lambda_b \right] = i f_{abc}\lambda_c\, .
\end{eqnarray}

Now we proceed to make a dimensional reduction of QCD, at the
classical level, from $4D$ to $2D$. For this we consider the
coordinates $x_{2,3}$ being compactified in a 2-Torus,
respectively the fields being periodic on the intervals ($0\leq
x_{2,3}\leq L=2\pi R$). Next we assume $L$ to be small enough in
order to get an effective model in $1+1$ dimensions.

Following this scheme we expand the fields in Fourier modes on the
compactified dimensions:

\begin{equation}
\label{com}
 G_\mu(x_0,x_1,x_2,x_3)=\sum_{n,m=-\infty}^\infty exp
\left\{\frac{2\pi i}{L}[n x_2+ m x_3]\right\}
G^{n,m}_\mu(x_0,x_1).
\end{equation}

By integrating heavy K-K modes Eq.~(\ref{com}) with $n,m\not=0$ in
the QCD generating functional
we obtain the low-energy effective
action for zero-modes which can be efficiently prepared along the Wilson
construction with the help of a cutoff $\Lambda$.
Accordingly, let us adopt the low-energy limit $|p_{0,1}|
\ll m_{KK}^{n,m}=\frac{2\pi}{L}\sqrt{n^2+m^2};\quad
n,m\not=0$. Hence if we choose
a $2D$ cutoff $\Lambda\simeq\frac{1}{R}=\frac{2\pi}{L}$ larger than the
typical momenta $p_{0,1}$, we are led to analyze the zero Fourier
modes only. Thus all relevant fields on the uncompactified $1 + 1$
space-time consist of zero modes.

To make contact with \mbox{'t Hooft} solution, we introduce light-cone
coordinates in the uncompactified space-time
$x_\pm=\frac{1}{\sqrt{2}}(x_0\pm x_1)$, then we obtain $A_\pm
=\frac{1}{\sqrt{2}}(G_0\pm G_1)$ and two scalar fields defined as
$\phi_{1,2}=G_{2,3}$.


By keeping only the zero K-K modes , we get the following
effective action:
\[
S=\int d^2x \frac{L^2}{{\tilde g}^2}\, tr\!\left[ (\partial_-
A_+-\partial_+ A_- - i[A_-,A_+])^2 + 2 (\partial_-\phi_1 - i
[A_-,\phi_1]) (\partial_+\phi_1 - i [A_+,\phi_1])+ \right.
\]
\begin{equation} \left.
2 (\partial_-\phi_2 - i [A_-,\phi_2]) (\partial_+\phi_2 - i
[A_+,\phi_2])+ [\phi_1,\phi_2]^2 \right] . \label{gluon1}
\end{equation}

Following the t'Hooft scheme we choose
the light-cone gauge $A_-=0$. Then the action
is transformed to:

\[
S=\int d^2x \frac{L^2}{{\tilde g}^2}\, tr\!\left[ (\partial_-
A_+)^2 - 2 i
A_+\left([\phi_1,\partial_-\phi_1]+[\phi_2,\partial_-\phi_2]\right)\right.
\]
\begin{equation} \left.
+
2(\partial_-\phi_1\partial_+\phi_1+\partial_-\phi_2\partial_+\phi_2)
+[\phi_1,\phi_2]^2\right] . \label{gluon2}
\end{equation}

This action is quadratic in $A_+$. To analyze the mass
spectra for fields $\phi_{1,2}$, we integrate over  $A_+$:

\[
 S_\phi=\int d^2x \,\, \frac{L^2}{{\tilde g}^2} tr\{
2\partial_-\phi_k\partial_+\phi_k + [\phi_1,\phi_2]^2 \]
\begin{equation}
 - \int d^2y
\left([\phi_1,\partial_-\phi_1]+[\phi_2,\partial_-\phi_2]\right)
\langle x|P.V.\frac{1}{\partial_-^2}|y \rangle
(\phi_k\partial_-\phi_k) \} .
\end{equation}

We expect the infrared mass generation for the two-dimensional
scalar gluons $\phi_{1,2}$ \cite{Coleman1}. Let us calculate it to
the leading order of perturbation theory. Applying the Background
Field Method we split $\phi=\frac{{\tilde g}}{L}(\phi_c+\eta)$,
with $\phi_c$ being a constant background. Then the one-loop
contribution is given by:

\begin{equation}
S^{(2)}=\int d^2x \,\mbox{\rm tr} \Bigl\{2\partial_-\eta_k\partial_+\eta_k
+\frac{{\tilde g}^2}{L^2}([\phi_{2c},\eta_1 ]^2 +
[\phi_{1c},\eta_2 ]^2)
\end{equation}
\[
-\frac{{\tilde g}^2}{L^2}\int d^2y \,
 \partial_-([\phi_{kc},\eta_k(x)]) \langle x|
\frac{1}{\partial_-^2}|y \rangle
\partial_-([\phi_{kc},\eta_k(y)])
\]
\[
+ 2\frac{{\tilde g}^2}{L^2}([\eta_1,\eta_2][\phi_{1c},\phi_{2c}]+
[\phi_{1c},\eta_2] [\eta_1,\phi_{2c}])\Bigl\} .
\]

The masses can be calculated from the quadratic part of the
effective potential,  after integration by parts:

\[
 S^{(2)}=\int d^2x \,\mbox{\rm tr} \Bigl\{ 2
\partial_-\eta_1\partial_+\eta_1+ 2 \partial_-
\eta_2\partial_+\eta_2
\]
\begin{equation} +\frac{{\tilde
g}^2}{L^2} ([\phi_{1c},\eta_1 ]^2 + [\phi_{2c},\eta_1 ]^2 +
[\phi_{1c},\eta_2 ]^2 + [\phi_{2c},\eta_2 ]^2 ) \Bigr\},
\end{equation}
They receive equal pieces from the two-dimensional gluon exchange and
the tadpole.

To estimate the masses of scalar gluons $\phi$ we use the Schwinger-Dyson
equations as self-consistency conditions:

\begin{equation}
 M^2=\frac{2 N_c {\tilde g}^2}{L^2} \int^\Lambda \!\!
\frac{d^2p}{(2\pi)^2} \,\frac{1}{p^2+M^2} \,
 = \frac{N_c {\tilde g}^2\,\Lambda^2}{8\pi^3}\,\,
 \log\frac{\Lambda^2+M^2}{M^2}\,, \label{glumass}
\end{equation}
thus $M^2$ brings an infrared cutoff as expected. We notice that
the gluon mass remains finite in the large-$N_c$ limit if the QCD coupling
constant decreases as $1/N_c$ in line with the perturbative law of
 $4D$ QCD. Respectively such a mass would dominate over the mass induced by
the Hosotani mechanism as the latter one is sub-leading in $1/N_c$ if being
provided by fermions in the fundamental representation \cite{hoso}.

We adopt the approximation $M \ll \Lambda\simeq 1/R$ to protect the low-energy
sector of the model and consider the
momenta $|p_{0,1}| \sim M$. Thereby we retain only leading terms
in the expansion in $\frac{p^2}{\Lambda^2}$ and
$\frac{M^2}{\Lambda^2}$, and also neglect the effects of the heavy
K-K modes in the low-energy Wilson action.

Now we proceed to the evaluation of the model (\ref{gluon1}) as a
generalization of the  \mbox{'t Hooft} $QCD_2$ and take a
bounded value of $M$. Then  the low-energy approximation
corresponds to the small coupling constant:

\begin{equation}
{\tilde g}^2 \sim \frac{M^2}{N_c \Lambda^2}\frac{1}{\log
\frac{\Lambda^2}{M^2}} \leq 1,
\end{equation}
in accordance to Eq.~(\ref{glumass}).
Thus the relevant coupling constant for the model  (\ref{gluon1}) is:

\begin{equation}
\label{G}
g^2= \frac{N_c\,{\tilde g}^2}{L^2}=\frac{N_c\,{\tilde
g}^2\,\Lambda^2}{4\pi^2} \sim \frac{M^2}{log
\frac{\Lambda^2}{M^2}}.
\end{equation}
This dimensional constant governs the meson mass spectrum of the
\mbox{'t Hooft} $QCD_2$ \cite{thooft0} and can be associated to a
particular meson mass, {\it i.e.} be fixed. Then the gluon mass is slowly
growing with the increasing cutoff $\Lambda$ and
and the "heavy-scalar'' expansion parameter is:
\begin{equation}
\label{A} A=\frac{g^2}{2\pi\,M^2}=\frac{1}{log
\frac{\Lambda^2}{M^2}}.
\end{equation}

 We observe that the limit $M \ll \Lambda$
supports consistently both the fast decoupling of the heavy K-K
modes and moderate decoupling of scalar gluons, the latter giving
an effective four-fermion interaction different from
\cite{Burkardt}. Nevertheless we prefer to keep first the gluon mass
finite and further on derive the expansion in parameter $A$ gradually from the
Bethe-Salpeter equation because for small  $A$ one recovers non-analytical
infra-red effects for the ground state wave function.
\subsection{Introducing fermions}
\label{C1fermions}

 Now let us make the K-K reduction of the quark part of
$QCD_4$ at the classical level. The original Lagrangian reads,

\begin{equation}
S = \int \! d^4 x {\bar \Psi}\,(i\gamma^\mu \, D_\mu - m)\,\Psi
\,,
\end{equation}
where the covariant derivative is defined conventionally: $D_\mu
= \partial_\mu -i\,G_\mu$.  The following representation of $\gamma$
matrices is employed:

\begin{eqnarray}
\gamma_0 = \left(\begin{array}{cc}\sigma_1& 0\\0& \sigma_1
\end{array}\right)\!,
              \gamma_1 =  \left(\begin{array}{cc}i\sigma_2& 0\\0& i\sigma_2
\end{array}\right)\!,
\gamma_2 =  \left(\begin{array}{cc}0 & i\sigma_3\\i\sigma_3 & 0
\end{array}\right)\!,
\gamma_3 =  \left(\begin{array}{cc}i\sigma_3 & 0\\0 & -i\sigma_3
\end{array}\right)\!,
\gamma_5 =  \left(\begin{array}{cc}0 & -i \sigma_3\\i\sigma_3 & 0
\end{array}\right)\!, \label{gamma}\nonumber
\end{eqnarray}
suitable for the purposes of K-K reduction. Then the bispinor is
decomposed into two-dimensional ones, $\Psi = \left(
\begin{array}{c}
\psi_a \\
\psi_b
\end{array} \right)$ with the help of the projectors,
$P_{a,b}=\frac12 (1 \pm
\gamma_2\,\gamma_5)$.

Now we compactify on the torus and expand the fermion fields in
Fourier modes of the compactified dimensions $x_{2,3}$:

\begin{equation}
\Psi (x_0,x_1,x_2,x_3) = \sum_{n,m=-\infty}^{+\infty} \exp
\left\{\frac{2i\pi}{L}(n\,x_2
+m\,x_3)\right\}\Psi^{n,m}(x_0,x_1)\,\, .
\end{equation}

As well as for gluon fields, we retain only the zero modes,
obtaining two spinors $\psi_a$ and $\psi_b$ of two components each
for the model in $(1+1)$ dimensions.

By keeping the zero K-K modes only, one arrives to the following effective
Lagrangian for the quark part of the model in $(1+1)$
dimensions, after a suitable rescaling of fermion fields:

\begin{eqnarray}
\label{lfermion}
 {\cal L} = i{\bar \psi}_a\left( \gamma^\mu \partial_\mu -
i\frac{g}{\sqrt{N}}\,\gamma^\mu \,A_\mu \right)\psi_a - m\,{\bar
\psi}_a\,\psi_a +
                               (a \rightarrow b) +\nonumber \\
\frac{g}{\sqrt{N}}\,\phi_1\left({\bar \psi}_a\,(i\sigma_3)\,\psi_b
+ {\bar \psi}_b\,(i\sigma_3)\,\psi_a\right) +
\frac{g}{\sqrt{N}}\,\phi_2\left({\bar \psi}_a\,(i\sigma_3)\,\psi_a
- {\bar \psi}_b\,(i\sigma_3)\,\psi_b\right)\,,
\end{eqnarray}
where we have used the definition for the $\gamma$ matrices in two
dimension: $\gamma^\mu = \left(\gamma^0,\gamma^1\right)$,
$\gamma^0 = \sigma_1$, $\gamma^1 = i\sigma_2$, inherited from
(\ref{gamma}).

The $\gamma$ matrices in the light cone basis are:
$\gamma^{\pm} = \frac{1}{\sqrt{2}}\left( \gamma^0 \pm
\gamma^1\right)$ and satisfy $\left\{\gamma^+,\gamma^- \right\} =
2$ and $\left(\gamma^+\right)^2 = \left(\gamma^-\right)^2 = 0$.

We consider the dressed quark propagator, which can be expressed
in term of one-particle irreducible self-energy $\chi$  as follows (see
fig.\ref{N1}):

\begin{equation}
S(p) = \frac{i}{\left[\gamma_-\,p_+ + \gamma_+\,p_- - m -
i\chi(p)\right]}\, .
\end{equation}

The equation for $\chi$ is given graphically in fig.\ref{N2}, in
the large $N_c$ limit and in the one-boson exchange approximation.
To solve the equation we decompose the matrix
$\chi$:

\begin{equation}
\label{chicom} \chi = \chi_+\,\gamma_- +  \chi_-\,\gamma_+ +
\chi_0 \, 1\!\textrm{l} \,,
\end{equation}
where $\chi_+,\chi_-$, and $\chi_0$ are functions of $p$. The
equations for the components (\ref{chicom}) of $\chi$   read:

\FIGURE[t]{\epsfig{file=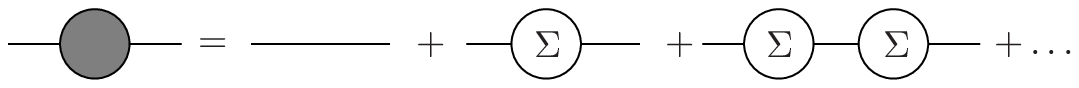,width=14cm} \caption{\label{N1}
{\sl The quark propagator in terms of the one-particle irreducible
self-energy where it is defined $\Sigma=i\chi$. }}}

\begin{equation}
 \chi_{+} = 2g^{2} \! \int \! \frac{d^{2}q}{(2
\pi)^{2}}\; \frac{ \left[ q -i\chi \right]_{-}}
      {2\left[q -i\chi \right]_{+}\left[ q -i\chi \right]_{-}
  - \left( m + i\chi_{0}\right)^{2} + i\epsilon}\;\,
                       \frac{{\bf P}}{(p - q)^{2}_{-}}
\end{equation}
\vskip .4cm
\[
\label{uno1} +\, 2g^{2} \! \int \! \frac{d^{2}q}{(2 \pi)^{2}}\,
\frac{1}{\left[2(p - q)_{+} (p - q)_{-} - M^{2} + i\epsilon
\right]}\, \frac{ \left[ q -i\chi \right]_{+}}
      {2\left[ q -i\chi \right]_{+}\left[ q -i\chi \right]_{-}
  - \left( m + i\chi_{0}\right)^{2} + i\epsilon} \, .
\]
\vskip .6cm

\begin{equation}
\label{dos} \chi_{-} = 2g^{2} \! \int \! \frac{d^{2}q}{(2
\pi)^{2}}\, \frac{1}{\left[2(p - q)_{+} (p - q)_{-} - M^{2} +
i\epsilon \right]}\, \frac{ \left[ q -i\chi \right]_{-}}
      {2\left[ q -i\chi \right]_{+}\left[ q -i\chi \right]_{-}
  - \left( m + i\chi_{0}\right)^{2} + i\epsilon}     \, .
\end{equation}
\vskip .6cm

\begin{equation}
\label{tress} \chi_{0} = 2g^{2} \!  \int \! \frac{d^{2}q}{(2
\pi)^{2}}\, \frac{-1}{\left[2(p - q)_{+} (p - q)_{-} - M^{2} +
i\epsilon \right]}\, \frac{ \left[ m + i\chi_{0} \right]}
      {2\left[ q -i\chi \right]_{+}\left[ q -i\chi \right]_{-}
  - \left( m + i\chi_{0}\right)^{2} + i\epsilon} \, .
\end{equation}
\vskip .6cm

We intent to solve these equation for $|p|\ll M$ and
at first sight one finds $\chi_+ =
\frac{i}{2\pi}\, \frac{g^{2}}{p_{-}}$, $\chi_-=0$, and $\chi_0 =0$
as solutions, that is, a \mbox{'t Hooft} solution for $\chi_+$ and
vanishing new terms. However it happens that this solution entails
imaginary masses for some meson states if one takes into account
the scalar field exchange, see appendix \ref{apendice1}.

On the other hand there exists another possibility, namely, if one
looks at Eq.(\ref{tress}) one could grasp that when the current
quark mass $m=0$, still a non-zero real solution for $\Sigma_0 = i
\chi_0$ exists, even in the case when $M^2 \rightarrow \infty$. To
explore this possibility, we start with $\chi_{-} = 0$ and a
constant $\chi_{0} \not= 0$, then in the zero  approximation in
$\frac{1}{M^{2}}$ expansion one arrives at the \mbox{'t Hooft} solution for
$\chi_{+}$:

\begin{equation}
\label{cinco} \chi_{+} = -i\pi{g}^{2} \! \int \! \frac{dq_{-}}{(2
\pi)^{2}}\;\,
 sgn(q_{-}) \frac{{\bf P}}{(p - q)^{2}_{-}}= \frac{i}{2\pi}\,
 \frac{g^{2}}{p_{-}} \,.
\end{equation}

To obtain $\chi_0$ in the chiral limit we treat Eq.(\ref{tress})
in the low-momentum limit ($M^2 \gg p^2$), then  the following
equation determines $\chi_0$:

\begin{equation}
\label{md1} 1 = 2g^{2} \!  \int \! \frac{d^{2}q}{(2 \pi)^{2}}\,
\frac{-i}{\left[q^2 - M^{2} + i\epsilon \right]}\, \frac{1}
{\left[ q^2 - m^2_d + i\epsilon \right]} =
\frac{g^2}{2\pi}\,\,\frac{1}{M^2 - m^2_d}\,\, \ln\!
\left(\frac{M^2}{m^2_d}\right) \, ,
\end{equation}
where we have defined $m^2_d = \Sigma_0 - \frac{g^2}{\pi}$. If one
assumes $M^2 \gg m^2_d$ one can solve the equation analytically to
find $\Sigma_0$:

\begin{equation}
\label{md2} \left. \Sigma^{2}_{0}\simeq \frac{g^2}{\pi} +
M^{2}\,\exp(-\frac{2\pi}{g^2}M^2) \simeq \frac{g^2}{\pi} +
\frac{M^4}{\Lambda^2} \right|_{\Lambda \rightarrow
\infty}\longrightarrow \frac{g^2}{\pi}\,.
\end{equation}

Let us notice that in virtue of Eq.(\ref{G}) in this limit
$\Sigma_0$ does not tend to zero but becomes a constant
$\frac{g^2}{\pi}$ which fully compensates the tachyonic quark mass
of the t'Hooft solution. We see also that the solution
Eq.(\ref{md2}) fully justifies the assumption $M^2 \gg m^2_d$.
Thus in our approximation $m^2_d =0$ in the chiral limit.

The quark propagator on this solution takes the following structure:

\begin{equation}
S(p) = \frac{ i\left( \gamma_{-}\left[ p -i\chi \right]_{+} +
\gamma_{+}\left[ p -i\chi \right]_{-} + m +i\chi_{0} \right)} {2(p
- i\chi)_{+} (p - i\chi)_{-} - (m + i\chi_{0})^{2} + i\epsilon} \simeq
i\frac{ \gamma_{-}\left[ {p}_{+} +
                          \frac{g^{2}}{2\pi {p}_{-}} \right]
                          + \gamma_{+}\, {p}_{-} + \Sigma_{0} }
  {{p}^{2} - m^{2}_d + i\epsilon}\, .
\end{equation}

Notice that the dynamical mass in the propagator is real and we don't have
tachyonic quarks as in \cite{thooft1}. The existence of tachyonic
quarks had been interpreted as a signal of confinement, but this is
incorrect as it was first understood in \cite{gross1,einhorn,Coleman}.

\FIGURE[t]{\epsfig{file=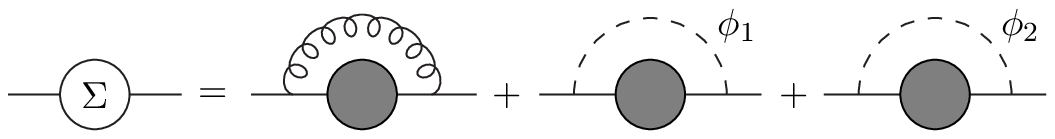,width=13cm}
\caption{\label{N2}{\sl The equation for $\Sigma$.}}}


\section{Mesons}
\label{C2mesons}
\subsection{The Bethe-Salpeter equation}

Now we want to study $q\,{\bar q}$ bound states. In our reduction we have four
possible combinations of quark bilinears to describe these states with
valence quarks:
$(\psi_a\,{\bar
\psi}_a),(\psi_a\,{\bar \psi}_b),$ $ (\psi_b\,{\bar
\psi}_a),(\psi_b\,{\bar \psi}_b )$. Each of these bilinears can be, of course,
supplemented with a number of gluons and adjoint scalars. Moreover, unlike
the t'Hooft model, the interaction
in our dimensionally reduced model mixes all sectors of the Fock space
with quark, antiquark and
any number of scalars, even in the planar limit,
as it has been realized before
in  the light-cone quantization approach \cite{light}.
However in the low-energy region for  $M^2 \gg m^2_d$ one can successfully
analyze the deviation from
the t'Hooft bound state wave function just retaining
the ladder  exchange both by gluon and by scalars for calculation of
the valence quark wave function and neglecting
vertex and self-energy corrections as well as the mixing with hybrid states
which are (superficially) estimated
to be of higher order in the  $1/M^2$ expansion. In this approach
the $q\bar q$ wave functions
satisfy the Bethe-Salpeter equation, for example $\psi_a\,{\bar
\psi}_a$ \mbox{satisfies} an equation given graphically in
fig.\ref{N3}. The Bethe-Salpeter equations for
these four states mix them (see fig.\ref{N3}) but if we define
the following combinations the equations decouple:

\begin{eqnarray}
\label{bstates1} \Lambda(p,q) = \frac{1}{2}\left( \Psi_{ab} -
\Psi_{ba} \right)\,, & &
\Delta(p,q) = \frac{1}{2}\left( \Psi_{aa} + \Psi_{bb} \right)\,, \\
\label{bstates2} \Theta(p,q) = \frac{1}{2}\left( \Psi_{ab} +
\Psi_{ba} \right)\,,  & & \Omega(p,q) = \frac{1}{2}\left(
\Psi_{aa} - \Psi_{bb} \right)\,.
\end{eqnarray}

These combinations have a direct interpretation in term of Dirac
bilinears of the theory in $(3+1)$ dimensions. In particular, the
scalar and pseudoscalar densities, which are within our scope in
this paper, can be represented by,

\begin{eqnarray}
\label{3+1a}
S = {\bar \Psi}\,\Psi = 2\,Tr(\Delta)\,, \\
\label{3+1b}
 P =  {\bar \Psi}\,\gamma^5 \,\Psi = -2i\,Tr(\Lambda\,
\sigma_3)\,,
\end{eqnarray}
were $\Psi$ is a four dimensional bispinor.

Respectively the manifest form of the Bethe-Salpeter equations on
wave functions
(\ref{bstates1}) is:

\[
\Lambda(p,q) = iS(p - q)\left( ig\gamma_{-} \right) \! \int \!
\frac{d^{2}k}{(2 \pi)^{2}}\; \left[\frac{{\bf P}}{(k -
q)_{-}^{2}}\,\, \Lambda(p, k) \right] \left( ig\gamma_{-} \right)
S(-q)
\]
\begin{equation}
\label{tres} -\, 2iS(p - q)\left( g \sigma_{3} \right) \! \int \!
\frac{d^{2}k}{(2 \pi)^{2}}\; \left[ \frac{1}{\left[2(p - q)_{+} (p
- q)_{-} - M^{2} + i\epsilon \right]}\,\, \Lambda(p, k) \right]
\left( g \sigma_{3} \right) S(-q)  \, .
\end{equation}
\vskip .6cm

\[
\Delta(p,q)= iS(p - q)\left( ig\gamma_{-} \right) \! \int \!
\frac{d^{2}k}{(2 \pi)^{2}}\; \left[\frac{{\bf P}}{(k -
q)_{-}^{2}}\,\, \Delta(p, k) \right] \left( ig\gamma_{-} \right)
S(-q)
\]
\begin{equation}
\label{cuatro} +\, 2iS(p - q)\left( g \sigma_{3} \right) \! \int
\! \frac{d^{2}k}{(2 \pi)^{2}}\; \left[ \frac{1}{\left[2(p - q)_{+}
(p - q)_{-} - M^{2} + i\epsilon \right]}\,\, \Delta(p, k) \right]
\left( g \sigma_{3} \right) S(-q) \, .
\end{equation}
\vskip 1.2cm
Further on we are going to focus our
analysis on
these equations to describe  scalar and pseudoscalar meson states.

\FIGURE[t]{\epsfig{file=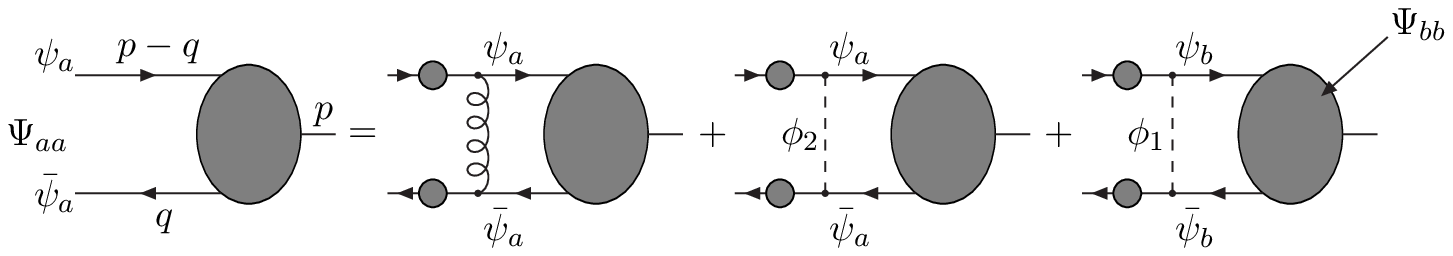,width=14cm}
\caption{\label{N3}{\sl Bethe-Salpeter equation for the state
$\Psi_{aa}$. Notice that in this equation the state $\Psi_{bb}$
appears.}}}

\subsection{Integral Equation for $\Lambda$ and Pseudoscalar Mesons}

Let us explore solutions of the matrix integral equation
(\ref{tres}). To do that we decompose the matrix $\Lambda$ as
follows:

\begin{equation}
\label{dec}
 \Lambda(p,q) = \Lambda_+\,\gamma_+ +
\Lambda_-\,\gamma_- + \Lambda_{+-}\,\gamma_+\,\gamma_- +
\Lambda_{-+}\,\gamma_-\,\gamma_+\,,
\end{equation}
then in components Eq.(\ref{tres}) yields four integral equations:

\begin{equation}
\label{A1} \Lambda_{+}(p, q) = \frac{-4i{g}^{2}(p - q)_{-} q_{-}}
{D[p-q]\, D[q]} \int \! \frac{d^{2}k}{(2 \pi)^{2}}
\left[\frac{{\bf P}}{(k - q)_{-}^{2}}\,\, \Lambda_{+}(p, k)
\right]
\end{equation}

\[
 +\, \frac{2i{g}^{2}}{D[p-q]\, D[q]}
 \! \int \! \frac{d^{2}k}{(2 \pi)^{2}}
\left[ \frac{2(p - q)_{-}q_{-}\Lambda_{-} - 2q_{-}\, \Sigma_{0}\,
\Lambda_{+-} + 2\, \Sigma_{0}(p -q)_{-}\Lambda_{-+} -
{\Sigma_{0}}^{2} \Lambda_{+} } {\left[ (k - q)^{2} - M^{2} +
i\epsilon \right] } \right]  \, .
\]

\begin{equation}
\label{A2} \Lambda_{-}(p, q) = \frac{2i{g}^{2}\, {\Sigma}^{2}_{0}}
{D[p-q]\, D[q]} \int \! \frac{d^{2}k}{(2 \pi)^{2}}
\left[\frac{{\bf P}}{(k - q)_{-}^{2}}\,\, \Lambda_{+}(p, k)
\right]
\end{equation}

\[
 +\,\frac{2i{g}^{2}}{D[p-q]\, D[q]}
 \! \int \! \frac{d^{2}k}{(2 \pi)^{2}}
\left[ \frac{ 2[p - q][q]\Lambda_{+} - 2[q]\, \Sigma_{0}\,
\Lambda_{-+} + 2\, \Sigma_{0}\,[p -q]\Lambda_{+-} -
{\Sigma_{0}}^{2} \Lambda_{-} } {\left[ (k - q)^{2} - M^{2} +
i\epsilon \right] } \right]  \, .
\]

\begin{equation}
\label{A3} \Lambda_{+-}(p, q) = \frac{2i{g}^{2}\, {\Sigma}_{0}\,(p
- q)_{-}} {D[p-q]\, D[q]} \int \! \frac{d^{2}k}{(2 \pi)^{2}}
\left[\frac{{\bf P}}{(k - q)_{-}^{2}}\,\, \Lambda_{+}(p, k)
\right]
\end{equation}

\[
 +\, \frac{2i{g}^{2}}{D[p-q]\, D[q]}
 \! \int \! \frac{d^{2}k}{(2 \pi)^{2}}
\left[ \frac{2(p - q)_{-}[-q]\Lambda_{-+} + [q]\, \Sigma_{0}\,
\Lambda_{+} - \Sigma_{0}\,(p - q)_{-}\Lambda_{-} +
{\Sigma_{0}}^{2} \Lambda_{+-}} {\left[ (k - q)^{2} - M^{2} +
i\epsilon \right] } \right] \, .
\]

\begin{equation}
\label{A4} \Lambda_{-+}(p, q) = -\frac{2i{g}^{2}\,
{\Sigma}_{0}\,\,q_{-}} {D[p-q]\, D[q]} \int \! \frac{d^{2}k}{(2
\pi)^{2}} \left[\frac{{\bf P}}{(k - q)_{-}^{2}}\,\, \Lambda_{+}(p,
k) \right]
\end{equation}

\[
 +\, \frac{2i{g}^{2}}{D[p-q]\, D[q]}
 \! \int \! \frac{d^{2}k}{(2 \pi)^{2}}
\left[ \frac{-2[p - q]q_{-}\Lambda_{+-} + q_{-}\, \Sigma_{0}\,
\Lambda_{-} - \Sigma_{0}\,[p - q]\Lambda_{+} + {\Sigma_{0}}^{2}
\Lambda_{-+} } {\left[ (k - q)^{2} - M^{2} + i\epsilon \right] }
\right]  \, ,
\]

where we have defined $D[q]= q^2-m^2_d+i\epsilon$, and $[q]=q_+ +
\frac{g^2}{2\pi}\left(\frac{{\bf P}}{q_-}\right)$.

In order to have control over the expansion in $\frac{1}{M^2}$ let us
introduce the following variables:

\begin{eqnarray}
{\bar \Lambda}_-(p,q)=
-2\frac{(p-q)_-\,q_-}{\Sigma^2_0}\,\Lambda_-(p,q) +
\Lambda_+(p,q)\,, &\;\;& {\bar
\Lambda}_{+-}(p,q)=-2\frac{q_-}{\Sigma^2_0}\,\Lambda_{+-}(p,q)
+ \Lambda_+(p,q)\,, \nonumber \\
{\bar
\Lambda}_{-+}(p,q)=2\frac{(p-q)_-}{\Sigma^2_0}\,\Lambda_{-+}(p,q)
+ \Lambda_+(p,q)\,.
\end{eqnarray}

One can see that in the large  $M^2$ limit the equations for these new variables contain only
terms, superficially proportional to $\frac{1}{M^2}$.
Meantime the
equation for $\Lambda_+$ contains \mbox{'t Hooft} operator
(which is of zero order in
$\frac{1}{M^2}$) as well as the terms of order $\frac{1}{M^2}$
in which
both the new variables and also $\Lambda_+$ appear. Since
the corrections to the \mbox{'t Hooft} equation for $\Lambda_+$ are  examined here at
the leading
order in $\frac{1}{M^2}$ expansion, we retain in this equation only the
terms including $\Lambda_+$:

\begin{equation}
\label{lambdaec1}
\Lambda_+(p,q)=2ig^2\,\frac{(p-q)_-\,q_-}{D[p-q]\,D[q]}\left[
-2\int \! \frac{d^{2}k}{(2 \pi)^{2}} \frac{{\bf P}}{(k -
q)_{-}^{2}}\,\, \Lambda_{+}(p, k)\right.
\end{equation}

\[ -\Sigma_0^2 \int \! \frac{d^{2}k}{(2 \pi)^{2}}
\frac{1}{(p-k)_-\,k_-}\,\frac{\Lambda_+(k)}{\left[ (k - q)^{2} -
M^{2} + i\epsilon \right] }
 -\frac{\Sigma_0^2}{(p-q)_-\,q_-} \int
\! \frac{d^{2}k}{(2 \pi)^{2}} \,\frac{\Lambda_+(k)}{\left[ (k -
q)^{2} - M^{2} + i\epsilon \right] }
\]

\[ \left.
+\, \frac{\Sigma^2_0}{(p-q)_-}\int \! \frac{d^{2}k}{(2 \pi)^{2}}
\frac{1}{k_-}\,\frac{\Lambda_+(k)}{\left[ (k - q)^{2} - M^{2} +
i\epsilon \right] } +\, \frac{\Sigma^2_0}{q_-}\int \!
\frac{d^{2}k}{(2 \pi)^{2}}
\frac{1}{(p-k)_-}\,\frac{\Lambda_+(k)}{\left[ (k - q)^{2} - M^{2}
+ i\epsilon \right] }\right].
\]

In the region $(k-q)^2 \ll M^2$ one can integrate over
$q_+$ on both sides of equation (\ref{lambdaec1}), obtaining the
following integral equation for the eigenvalues $p^2$:

\[
\left[ p^2 - \frac{m^2_d}{x} - \frac{m^2_d}{(1 - x)} \right]
\phi(x) = \, -\frac{g^2}{\pi}\, \int^{1}_{0} \! dy \, \frac{{\bf
P}}{(x - y)^{2}}\,\, \phi(y) -\,
A\,\frac{\Sigma^{2}_{0}}{x^{1-\beta}} \int^{1}_{0} \! dy  \,
\frac{\phi(y)}{(1 - y)^{1-\beta}}
\]

\begin{equation}
\label{n1}   -\, A\,\frac{\Sigma^{2}_{0}}{(1 - x)^{1-\beta}}
\int^{1}_{0} \! dy  \, \frac{\phi(y)}{y^{1-\beta}} +\,
A\,\frac{\Sigma^{2}_{0}}{[(1 - x)\,x]^{1-\beta}} \int^{1}_{0} \!
dy \, \phi(y) +\, A\,\Sigma^{2}_{0}\, \int^{1}_{0} \! dy  \,
\frac{\phi(y)}{[(1 - y)\,y]^{1-\beta}}  \, .
\end{equation}

Where we have conventionally introduced $\phi(q_-)=\int
dq_+\,\Lambda_+$ and dimensionless variables $q_-=xp_-$,
$k_-=yp_-$ as well as the expansion parameter $A$, previously
defined in Eq.(\ref{A}). The integral equation (\ref{n1}) has a
symmetric kernel that guarantees that the eigenvalues $p^2$ are
real. It is also symmetric  under the transformation $x\rightarrow
(1-x)$ found by \mbox{'t Hooft} in \cite{thooft1}.

To the first order in $A$ the straightforward  calculation from
Eq.(\ref{lambdaec1}) gives $\beta=0$, but such a kernel makes this
equation too singular to have any reasonable solutions. This
singularity must be softened by re-summation of a series of similar
singularities of higher order in $A$. Just in order to simulate
this re-summation   we have introduced $\beta$ which effectively
regularize the integral equation. The coefficient $\beta$ governs
the behavior of the solution of Eq.(\ref{n1}) near the end points.
We fix it by imposing self consistency of solutions near the
boundaries. Although the parameter $\beta$ is small and vanishing
with $A \rightarrow 0$, its presence guarantees the positivity of
the spectrum (see appendix \ref{apendice2}).


\section{Integral Equation and Approximate Spectrum}
\label{C3integral}

To explore solutions to Eq.(\ref{n1}) we examine small $A$, which
allows us to put $m^2_d =0$. Working in this limit we select out
consistently the wave functions that behave as $x^\beta$ or
$(1-x)^\beta$ on the boundaries, then $\beta$ is fixed  to
$\frac{A}{2}$ in order to cancel the principal divergences,
$x^{\beta-1}$ and $(1-x)^{\beta -1}$, that appear when $x$ is near
the end points.

Evidently Eq.(\ref{n1}) does not mix even and odd functions with
respect to the symmetry $x \Longleftrightarrow 1 - x$. On the
other hand the ground state should be an even function. When
inspecting the wave function end-point asymptotics from the
integral equation (\ref{n1}) one derives the following even
function as a  ground state solution for $A \rightarrow 0$ limit:
\begin{equation}
\label{pion}
\phi_0(x)=(4x[1-x])^{\frac{A}{2}}-\frac{1}{\pi}\,(4x[1-x])^{\frac{1}{2}}\,.
\end{equation}

This is basically a non-perturbative result that differs from 't
Hooft solution in the $A \rightarrow 0$ limit, giving $p^2=0$, as
we would expect from spontaneous chiral symmetry breaking in $4D$.
Notice that the second term in Eq.(\ref{pion}) is a leading
contribution to the ground state solution for \mbox{'t Hooft} equation
with $m^2=\frac{g^2}{\pi}$ \cite{thooft1}, and the first term
compensates powers of $A$ when we insert $\phi_0(x)$ in
Eq.(\ref{n1}) and take the $A \rightarrow 0$ limit. The entire
result is a shift of the ``unperturbed'' ground state
$p^2=7.25\frac{g^2}{\pi}$ (just the \mbox{'t Hooft} solution with
$m^2=\frac{g^2}{\pi}$) to the non-perturbative solution with
$p^2=0$ in the $A \rightarrow 0$ limit. For the other massive
states we are unable to find analytic solutions, as happens with
\mbox{'t Hooft} equation, but we could estimate them working with the
Hamiltonian matrix elements $p^2(\phi,\phi)=(\phi,H\phi)$ and
using the regular perturbation theory, starting from \mbox{'t Hooft}
solutions ($A=0$) \cite{thooft1}. Formally the set of solutions to
Eq.(\ref{n1}) is composed of even and odd functions  with respect
to the reflection $x \Longleftrightarrow 1 - x$, but only the even
functions have physical meaning as describing pseudoscalar states.
It occurs because the $4D$ pseudoscalar density could be written
as a combination of $2D$ pseudoscalar densities for which it is
possible to use the arguments given in \cite{gross1}. Thus for the
pseudoscalar sector only the set of even solutions of
Eq.(\ref{n1}) is selected out. Similarly for scalar states the
equation differs from Eq.(\ref{n1}) by signs of perturbation
terms, {\it i.e.} $A \longrightarrow - A$ and the appropriate set
of solutions consists of odd functions, then only perturbative
massive solutions are possible for these states. As we have significantly
departed from $QCD_4$ neglecting heavy K-K states and taking the limit
 $A \rightarrow 0$ we don't  make any numerical fits to some observable
mesons with quantum numbers of scalar and pseudoscalar radial excitations
just postponing this work till a more systematic estimations of neglected
terms.

\section{Conclusions}
\label{C4conclusion}

We summarize our points. Quantum Chromodynamics at low energies
has been decomposed by means of dimensional reduction from $(3+1)$ to
$(1+1)$ dimensions and  a low energy effective model in $2D$ has
been derived. It is presumably more realistic than just taking
QCD in $2D$ because the model includes, in a nontrivial way,
physical (transverse) gluon degrees of freedom. We treat this
model in the equal-time quantization approach using
Schwinger-Dyson equations instead of Hamiltonian quantization, and
we argued that the adjoint scalar fields dynamically gain  masses. We did
an explicit analysis of the meson bound states by using
Bethe-Salpeter equation in the limit of large scalar masses.

We found that the model has two regimes classified by the
solutions of quark self-energy. In one case, the perturbative one,
we have tachyonic quarks in the chiral limit as in \cite{thooft1}
and our model could be interpreted as a perturbation from the
result of chiral QCD  in $2D$. But we are not allowed to consider
that possibility because the spectrum for the lowest $q {\bar q}$
bound states becomes imaginary once we consider small but nonzero
$A$, see appendix \ref{apendice1}. This means that the \mbox{'t Hooft}
chiral solution is not a stable background to start a
perturbation to $QCD_2$ that includes transverse degrees of
freedom. Also in this regime the equations that describe the mass
spectrum for scalars and pseudoscalars are equal, certainly we
have oversimplified something. On the other hand the non-perturbative
solution to quark self-energy supports non-tachyonic quarks with
masses going to zero, in the chiral limit.

We focused our attention on bound states of quark-antiquark in the
valence quark approximation which is right for low energies.
It was shown that $q{\bar q}$ states
constructed in our reduced model have a direct interpretation in
terms of their  properties under Lorentz transformations inherited from
$(3+1)$ dimensions. Respectively,
we have different equations that govern the scalar and
pseudoscalar mass spectrum and in both cases the spectrum is
positive definite with a massive ground state for scalars and a
massless ground state for pseudoscalars.

The pseudoscalar and scalar sectors of the theory have been
analyzed and a massless solution for the pseudoscalar ground state
has been found. We interpreted this solution as a ``pion'' of our
model although it is a quantum mechanical state only
existing in the large-$N_c$ limit. The appearance of massless boson in a two
dimensional theory in a particular limit is not a new
feature and has been studied in
\cite {witten,zhitnitsky,vento1} and more recently justified in
\cite{Faber}. In particular in the planar $QCD_2$ the \mbox{'t Hooft}
pion solution
is a massless boson bound state, related with a non-anomalous $U(1)$ symmetry.
We expect that our pion solution Eq.(\ref{pion}) has  more
physical properties from the $4D$ theory, as compared to the
massless solution of QCD $2D$. For the latter one, it has been
proved that the \mbox{'t Hooft} pion state is decoupled from the rest of
the states \cite{krauth}. On the contrary,
the argument of non-anomalous $U(1)$
symmetry that produced the decoupling of the massless scalar field
in $QCD_2$ does not hold in our model.

Indeed in $QCD_2$ the $U(1)$ vector current $J^\mu = {\bar
\psi}^a\gamma^\mu \psi^a + {\bar \psi}^b\gamma^\mu \psi^b$ is
conserved then $J^\mu =\epsilon^{\mu\,\nu}\partial_\nu \phi(x)$.
Due to $\gamma_\mu \,\gamma_5=\epsilon_{\mu\,\nu}\gamma^\nu$ we
can write the $U(1)$ axial current as follows
$J^5_\mu=\partial_\mu \phi(x)$, then if the axial current is also
conserved we have $\partial^\mu
\partial_\mu \phi =0$, a massless decoupled scalar field. This
scalar field is identified with the $\phi_0=1$ \mbox{'t Hooft} state.

 In our model the situation is different because
the $U(1)$ axial symmetry is explicitly broken by the interaction
term in Eq.(\ref{lfermion}), but our model has another symmetry
due to the breakdown of the $4D$ Lorentz symmetry $SO(1,3)$ to
$SO(1,1)\times O(2)$, where $O(2)$ now becomes an internal
symmetry. The vector and axial currents associated with this
symmetry are conserved and the axial symmetry corresponds
precisely to the projection of the $U(1)$ axial symmetry in $4D$
to our model in $2D$. Our massless pseudoscalar meson is related
with these currents, then we cannot expect the same behavior for
our massless pseudoscalar as in $QCD_2$ because we are talking
about different symmetries. For instance, the vector and axial
$O(2)$ currents do not satisfy the same relation as $J_\mu$ and
$J_\mu^5$ in $QCD_2$. On the other hand the possibility to have a
coupled massless scalar in the large-$N_c$ does not contradict the
Coleman theorem as it has been first pointed out in \cite{witten},
and particular in our case because it is not a quantum Goldstone
field but a quantum-mechanical state. We stress that this
composite massless state can be related to the real pion, to some
extent, at the tree level as the chiral $2D$ perturbation theory
being renormalizable is drastically different from the $QCD$ $4D$
Ch.P.T. \cite{gasser}.

\acknowledgments

We are grateful to A. Ivanov for useful comments on massless bosons in two
dimensions.
This work is partially supported (A.A.) by INTAS-2000 Grant (Project 587)
and the Program "Universities of Russia:
Basic Research" (Grant 02.01.016).
The work of P.L. is supported by a Conicyt Ph. D. fellowship(Beca Apoyo Tesis Doctoral).P.L. wishes to thank the warm hospitality extended to him during his visits to Istituto Nazionale
di Fisica Nucleare Sezione di Bologna and also to M. Cambiaso for many helpful discussions.
The work of J.A. is partially supported by
Fondecyt \# 1010967. He thanks the hospitality of LPTENS (Paris), Universidad de
Barcelona and Universidad Aut\'onoma de Madrid. P.L. and J.A. acknowledge financial
support from the Ecos(France)-Conicyt(Chile) project\# C01E05.

\appendix
\section{Quark-Self Energy, First Solution}
\label{apendice1} Let us consider the t'Hooft solution for the quark
self energy $\chi_+=\frac{i}{2\pi}\frac{g^2}{p_-}$, $\chi_0=0$ and
$\chi_-=0$ and explore what this solution implies for the bound
state spectrum of quarks and antiquarks.
We proceed in the same way as before
obtaining a similar Bethe-Salpeter (BS) equation for
$\Lambda(p,q)$ Eq.(\ref{tres}), but now the dynamic mass is the
one found in \cite{thooft1} ($m^2_d=m^2-\frac{g^2}{\pi}$) and
$\Sigma_0$ is zero. In the chiral limit
($m=0$), we write the BS equation in terms of the
decomposition given in Eq.(\ref{dec}) and obtain two decoupled
sets of equations: one for $\Lambda_{+}(p, q)$ and $\Lambda_{-}(p,
q)$ and the other one for $\Lambda_{+-}(p, q)$ and $\Lambda_{-+}(p,
q)$. To calculate corrections to the  \mbox{'t Hooft} equation
for $\Lambda_+$ we retain just the first set of equations:

\begin{equation}
\label{AP1} \Lambda_{+}(p, q) = \frac{-4i{g}^{2}(p - q)_{-} q_{-}}
{D[p-q]\, D[q]} \Bigg[
\end{equation}

\[
 \int \! \frac{d^{2}k}{(2 \pi)^{2}} \frac{{\bf P}}{(k -
q)_{-}^{2}}\,\, \Lambda_{+}(p, k)
 -\!\int
\! \frac{d^{2}k}{(2 \pi)^{2}} \frac{\Lambda_{-}(p,k)} {\left[ (k -
q)^{2} - M^{2} + i\epsilon \right] }\Bigg],
\]

\begin{eqnarray}
\label{AP2} \Lambda_{-}(p, q) = \frac{4i{g}^{2}\,[p -
q][q]}{D[p-q]\, D[q]} \! \int \! \frac{d^{2}k}{(2 \pi)^{2}} \frac{
\Lambda_{+}(p,k)} {\left[ (k - q)^{2} - M^{2} + i\epsilon
\right]},
\end{eqnarray}
where we can write $[q]=\frac{1}{2}(q^2-m^2_d)\left(\frac{{\bf
P}}{q_-}\right)$. Then the equation for $\Lambda_-(p,q)$ becomes:

\begin{eqnarray}
\label{AP3} \Lambda_{-}(p, k) = i{g}^{2}\,\left(\frac{{\bf
P}}{k_-}\right)\left(\frac{{\bf P}}{(p-k)_-}\right)\! \int \!
\frac{d^{2}k'}{(2 \pi)^{2}} \frac{ \Lambda_{+}(p,k')} {\left[ (k -
k')^{2} - M^{2} + i\epsilon \right]},
\end{eqnarray}
now we replace this equation into Eq.(\ref{AP1}) and integrate
over $k$ in the second term retaining only the first
$\frac{g^2}{M^2}$ order. Further on, it is possible to integrate
over $q_+$ at both sides of the equation obtaining the following
integral equation for $\phi(x)$:

\[
\left[ {\bar p}^2 + \frac{1}{x(1-x)}\right] \phi(x) =
-\int^{1}_{0} \! dy \, \frac{{\bf P}}{(x - y)^{2}}\,\, \phi(y)
\]

\begin{equation}
\label{AP4}
 +\,A \int^{1}_{0} \! dy \frac{{\bf P}}{(x - y)}\, \ln
\left( \frac{(1 - x)\,y}{(1 - y)\,x}\right) \phi(y),
\end{equation}
where we have defined ${\bar p}^2=\frac{\pi}{g^2}p^2$.
One might solve this equation
by using perturbation theory near \mbox{'t Hooft} solution ($A=0$). We
know from \cite{thooft1} that the ground state of \mbox{'t Hooft}
equation is $\phi_0(x)=1$ with ${\bar p}^2=0$, but the perturbation term in
Eq.(\ref{AP4}) generates a negative correction to ${\bar p}^2$.
Thereby for non-zero values of $A$ we obtain for the
ground state an imaginary mass.

\section{Positivity of the Spectrum}
\label{apendice2}

Here we show that the spectrum of integral equation
(\ref{n1}) is indeed positive as stated above. We know that the
lowest state should  be an even function. For any
function that goes to zero faster than $x^{\frac{A}{2}}$ or
$(1-x)^{\frac{A}{2}}$ when $x$ goes to zero or one, respectively,
we can consider the terms proportional to $A$ as perturbations to
\mbox{'t Hooft} term which is positive definite. Then the only possible
trial functions that could generate negative $p^2$ are nearly
constant functions. Let us explore this possibility and
consider a function $f(x)=[x(1-x)]^\gamma$, with $\gamma$ small
but arbitrary, then we multiply both sides of Eq.(\ref{n1}) by
$f(x)$ and integrate over $x$ obtaining (see appendix of reference
\cite{harada}):

\begin{equation}
\label{apx2} {\bar p}^2 \langle f(x)|f(x) \rangle
=2^{4\gamma-2}\,\gamma\,\frac{\Gamma[\gamma]^4}{\Gamma[2\gamma]^2}
-2A\pi\,\frac{\Gamma[\gamma+\beta]^2}{\Gamma[\frac{1}{2}+\gamma+\beta]^2}+2A\pi\,\frac{\Gamma[\gamma+\beta]\,\Gamma[1+
\gamma]}{\Gamma[\frac{1}{2}+\gamma+\beta]\,\Gamma[\frac{3}{2}+\gamma]}\,.
\end{equation}
If we assume $A$ and $\gamma \equiv \alpha A$ to be small, only the first two terms of the
right hand side of
Eq.(\ref{apx2}) are relevant. The divergent part of Eq.(\ref{apx2}) is:
\begin{equation}
{\bar p}^2 \langle f(x)|f(x) \rangle \sim \frac{1}{\gamma}\left[1-
\frac{2A\gamma}{(\gamma+\beta)^2}\right]=\frac{1}{\gamma}\left[1-
\frac{2\alpha}{(\alpha+\frac{1}{2})^2}\right].
\end{equation}
Then for any value of $\alpha$ the expression between brackets is evidently
positive.

Notice that the existence of a non-zero $\beta$ guarantees the
positivity of the spectrum, when $\gamma$ goes to zero and $A$ is
small but different from zero.



\begin{thebibliography}{References}
\bibitem{footnote1} A similar particle content in $2D$ was set in the
extended $QCD_2$ model treated under the light-cone quantization \cite{light}.
However this quantization leads to the trivial QCD vacuum and has difficulties
to describe such crucial nonperturbative phenomena
as infrared mass generation for transverse
gluons and  the spontaneous chiral symmetry breaking.
\end{thebibliography}
\end{document}